\documentclass[amsmath,amssymb,floatfix,preprint,showpacs]{revtex4}
\usepackage{amsfonts}
\usepackage{graphicx}
\usepackage{dcolumn}
\usepackage{bm}
\usepackage{hyperref} 
\usepackage{latexsym}
\usepackage{float}

\begin{document}
\title{Explosive Percolation in the Human Protein Homology Network}

\author{Hern\'an D. Rozenfeld}
\author{Lazaros K. Gallos}
\author{Hern\'an A. Makse}
\affiliation{Levich Institute and Physics Department, City College of New York,
New York, NY 10031, USA}

\begin{abstract}
We study the explosive character of the percolation transition in a
real-world network. We show that the emergence of a spanning cluster
in the Human Protein Homology Network (H-PHN) exhibits similar
features to an Achlioptas-type process and is markedly different from
regular random percolation. The underlying mechanism of this
transition can be described by slow-growing clusters that remain
isolated
until the later stages of the process, when the addition of a small number of
links leads to the rapid interconnection of these modules into a giant cluster.
Our results indicate that the evolutionary-based process that shapes
the topology of the H-PHN through
duplication-divergence events may occur in sudden steps, similarly to
what is seen in first-order phase transitions.
\end{abstract}

\pacs{
64.60.ah, 
64.60.aq,	
64.60.-i,	
87.18.Vf    
}
\maketitle

\section{INTRODUCTION}


Percolation is a heavily studied process which can provide critical information on the large-scale
connectivity of a system, and quantifies thus its ability to efficiently transfer information,
resources, etc~\cite{perco2, perco3, perco4}. A typical bond percolation process starts with an empty space, 
where random links are added continuously until the largest existing cluster spans the entire system.

An enormous number of variations have been reported in the literature, both for lattice percolation
and for network percolation, where links connect nodes that are not embedded in Euclidean space~\cite{dorogovtsev,cohen,boguna,rozenfeld,garas}.
Typically, the emergence of the spanning cluster follows a second-order phase transition.
This is measured as a steep, but continuous, increase in the number of nodes in the largest component.
In a recent paper, Achlioptas {\it et al.}~\cite{achlioptas} proposed a percolation procedure that exhibits a
first-order phase transition, seen as a discontinuity in the size of the largest component,
which was described as {\it explosive percolation}. The model is based on the product rule,
which favors the appearance and separation of small clusters that are progressively joined into
larger ones, so that the spanning cluster (a) emerges late in the process and (b) requires a relatively
small number of links that finally join the smaller isolated clusters. Much work has been done on 
delaying or accelerating the appearance of the largest component~\cite{Bohman01, Bohman06, Spencer07}.
Following the work of Achlioptas {\it et al.}, Ziff~\cite{ziff} showed that this process gives different results than 
random percolation in two-dimensional lattices, while Cho {\it et al.}~\cite{cho} and Radicchi {\it et al.}~\cite{radicchi} 
showed than in random scale-free networks a first-order transition can only be achieved with a degree 
distribution exponent $\gamma>3$.

In this work, we present evidence that this model is ideally suited for describing the
evolution of a modular network and can be also interpreted as a possible mechanism that explains the
growth process of a real network. Our results are based on a Protein Homology Network (PHN)~\cite{medini,SIMAP}. This network is composed of highly connected clusters of homologous nodes, while links between nodes of low homology generate  inter-cluster connections. This structure is similar to the presence of strong links within communities and weak links between communities, as suggested by Grannoveter~\cite{granovetter} for social systems.

\section{A DETERMINISTIC MODEL FOR A FIRST-ORDER PERCOLATION TRANSITION}
\label{deterministic}

The product rule (PR) for an Achlioptas process works as follows:
We start with $N$ disconnected nodes. At each step we randomly select two pairs of nodes as candidate links. Each link connects two clusters of size $S_A$ and $S_B$, respectively. Of the two links, we only add the one for which the product $S_AS_B$ is smaller and discard the other. We repeat this process until a giant component emerges. Clearly, under this rule it is more probable to connect two small clusters with each other, rather than large ones.

\begin{figure}
\includegraphics*[width=0.40\textwidth]{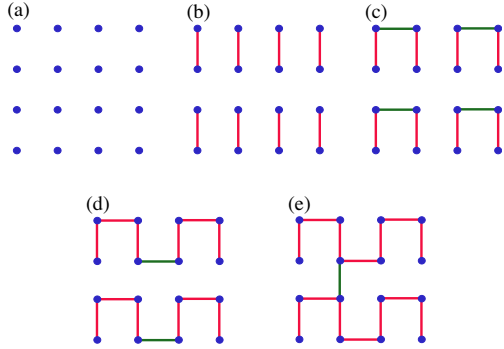}
\caption{Deterministic model leading to explosive percolation. In (a) we start with $N$ nodes without links.
In (b) we connect nodes in pairs. In (c) and (d) we iteratively connect each pair of clusters with one link to form larger components.
This mechanism continues recursively by joining smaller clusters into larger ones, and the spanning component
only emerges at the end of the process, when the entire network is connected, as in (e).
}
\label{model_net}
\end{figure}

In order to gain insight into the conditions under which we obtain a sharp first-order percolation transition, we can remove the element of randomness.
We present a simple deterministic process, inspired by the basic Achlioptas model, that also leads to explosive percolation,
and is doing so in an even more dramatic way than in the original PR model. At the
same time it highlights the basic requirements for explosive percolation: a large number of small clusters that grow at
equal pace until they are all joined together with the addition of very few links.
In the Achlioptas model, as explained above, one selects two random candidate links and accepts the one that minimizes the product of the component sizes. As a possible extension one may randomly select three pairs of nodes and select one according to the same PR rule. As a result, we would obtain a percolation transition that appears at a later stage compared to the basic Achlioptas model.
One may extend this idea to more candidate links and still select one link according to the PR rule. In the limit in which we select one link (according to the PR rule) among {\it all} possible pairs of nodes, we fully eliminate the stochastic aspect of the Achlioptas model and still preserve the first-order percolation transition.
This model is useful for understanding the conditions leading to a sharp transition, but obviously it cannot capture the complexity of the Achlioptas model, resulting from its stochastic aspect and the competition between links.
This model serves as a prototype that helps us to visualize the behavior of the percolation transition in a real-world complex network as we show later in Section~\ref{phn_section}.
The algorithm is as follows (see Fig.~\ref{model_net}):
At step $t=0$ we start with an empty network of $N=2^m$ nodes, for $m>0$. At step $t=1$ we add a link between any pair of nodes,
leading to a network of $N/2$ separate components. Following this, 
in every step $t$ we join a pair of two components from step $t-1$.

\begin{figure}
\includegraphics*[width=0.40\textwidth]{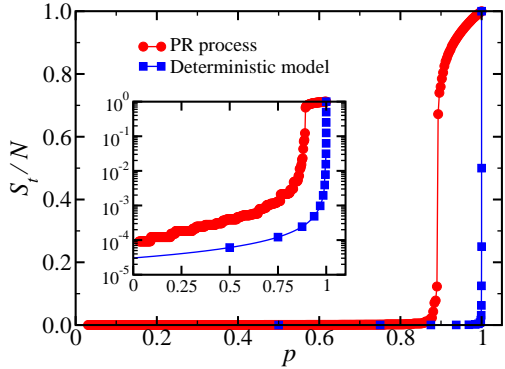}
\caption{Fraction of nodes in the largest component as a function of the added links percentage $p$. We compare the
results for the deterministic model with the PR process, showing the pronounced first-order
transition of the model. Both networks include $N=32768$ nodes.
Inset: The same plot with a logarithmic y-axis highlights the explosive character of the transition,
where the addition of the final 1\% of the links results in 3 orders of magnitude increase in the largest cluster.
}
\label{size_of_largest_components}
\end{figure}

Notice that according to this process, the number of components $n_t$ at step $t$ is always half the number of components of step $t-1$,
and therefore $n_t= 2^{m-t}$, for $0 \leq t < m$. Moreover, the number of nodes in each component
is $S_t=2^t$ (all components at a given step $t$ have the same number of nodes).
The number of new links at step $t$ is $n_{t-1}/2$, and therefore the total number of links in the network up to step $t$ is 
\begin{equation}
M_t = N \sum_{i=1}^{t} 2^{-i} = N(1-2^{-t}).
\end{equation}
The resulting network is a tree, so that the total number of possible links in the network is $M \equiv N-1$.
Consequently, the fraction of links added to the network up
to time t is $p \equiv M_t/M=M_t/(N-1) \approx 1-2^{-t}$, for $N >> 1$.
Therefore, the dependence of the largest component size $S_t$ on the fraction of added links $p$ follows
\begin{equation}
S_t = \frac{1}{1-p},
\end{equation}
which exhibits a singular point at $p=1$. This singularity is the hallmark of
a discontinuous or first-order percolation phase transition, similarly to the reported
result of Ref.~\cite{achlioptas}. In Fig.~\ref{size_of_largest_components} we show the size of the largest component as
a function of the fraction of links added to the network, and we compare the PR process to the above
presented deterministic model for the same network size. Both cases exhibit explosive percolation, but obviously
the deterministic process leads to a much sharper transition, even for the small finite size that is considered here ($N=32768$ nodes).
This model can be considered as an optimized (albeit trivial) explosive percolation process, since the transition occurs
at the latest possible stage, with a very delayed time of ``explosion'' and a very steep jump.

\section{PRODUCT RULE ON A REAL NETWORK}
\label{phn_section}

In this section we present a real-world complex network where the idea of explosive percolation
can be readily applied.
In the Appendix we describe how we built the Human Protein Homology Network (H-PHN),
which is a weighted network with the weights denoting the degree of similarity (homology) between
two proteins. Homologous proteins have been shown to organize themselves in network modules~\cite{medini}.
A large number of proteins may have evolved through duplication-divergence events from a
single ancestral protein, preserving thus the phylogenetic relationships in the network representation.
This, in turn, leads to a large clustering of homologous proteins in distinct families. 

The existence of weights in a modular network that has evolved with time provides an ideal
case of a network that can be explained on the basis of a PR model. Our hypothesis
is that, in general, there exists a spanning skeleton of the network which connects all the
different network areas. Locally, though, there are modules of well-connected proteins, and
it is much more probable that the links of new proteins have a much larger weight
within a module rather than with proteins that are further away. These dense
modules are then connected with each other through weaker links. In the terminology
of the PR model, this corresponds to an increased probability of connections between small clusters compared to the growth of 
already large clusters. 
Thus, if the above assumption holds and we only consider large weights we will be able to identify the dense modules.
As we lower the weight threshold, modules will merge into larger clusters until the entire network is connected.

\begin{figure}
\includegraphics*[width=0.40\textwidth]{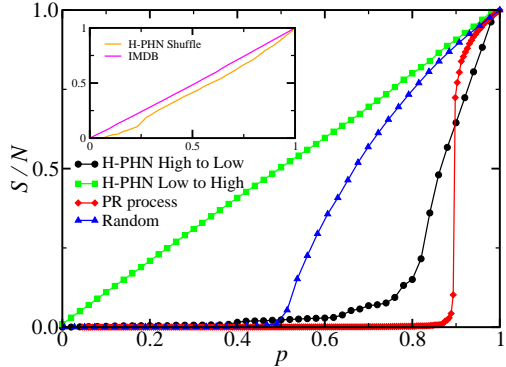}
\caption{Fraction of nodes in the largest cluster of H-PHN as we add a percentage $p$ of links according
to different rules: a) links are added in decreasing weight order, b) links are added in increasing
weight order, c) links are chosen according to the PR model, and d)
links are added randomly. 
Inset: Same quantity a) when the structure of H-PHN remains the same but link weights are randomly redistributed, and b)
for the weighted IMDB co-acting network.
}
\label{largest_component_phn}
\end{figure}

To study the idea of explosive percolation in the H-PHN we start with an empty network of all proteins in the H-PHN
and no links between them. Then, starting with the largest weight link, we add one link at a time in decreasing
order of the weight, which corresponds to the score ratio $SR$ (see Appendix). 
In order to directly compare the different models, we do not increase the value of $p$ when a link connects two nodes within the same cluster.  
%

This process initially leads to well-connected families of highly homologous proteins,
that are interconnected at later stages by links with smaller values of $SR$,
resembling the requirements for explosive percolation explained in Ref.~\cite{achlioptas} and in Section~\ref{deterministic}.
This behavior can be seen in the size of the largest component as links are added to the network
(Fig.~\ref{largest_component_phn}). Clearly, the largest cluster in the network remains very small
even when we have added a significant number of links, e.g. as much as 80\% of the total network links.
This small size is an indication that links are added locally, and up to a given weight there is not
a significantly large spanning cluster. The subsequent addition a relatively small number of links, causes the entire network
to become connected through the merging of the small clusters in one spanning entity.
This behavior is similar to the one observed when we employ a purely PR Achlioptas process for the same
number of nodes (without considering the H-PHN links). The product rule, as described in the
previous section, favors small clusters that increase slowly in size until the explosion stage
where the spanning cluster emerges, which can explain the similarities between the two curves in the
plot, despite the sharper transition of the PR model compared to the real data of H-PHN.

\begin{figure}
\includegraphics*[width=0.40\textwidth]{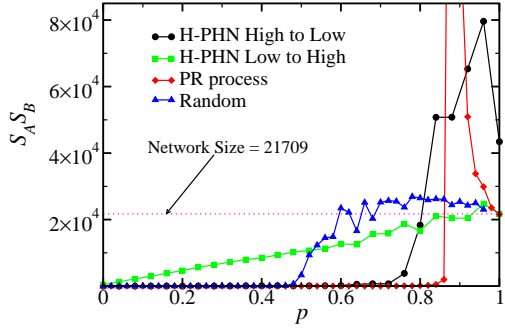}
\caption{Average product of the cluster sizes that are connected when we add a new link,
as a function of the fraction $p$ of links that have already been added. Symbols
are the same as in Fig.~\ref{largest_component_phn}. The horizontal line corresponds
to the network size.
}
\label{component_product}
\end{figure}

For comparison, we also plot the size of the largest component when links are added in increasing order of the $SR$ (low to high),
and also when they are added at random. When links are added in increasing order of the $SR$ (red points),
the percolation transition appears at a very early stage, and is practically non-existent. Links with small
values of $SR$ connect dissimilar proteins which, in general, do not belong to the same module.
Therefore, a link with a small value of $SR$ is equivalent to a long-range connection, which leads to a rapid growth
of the largest cluster and leaves all the other components small. This also explains the linear increase of $S$ with $p$, where almost every new link attaches itself to the largest cluster.
Random link addition has a milder effect, in accordance to the idea of percolation in random networks,
where after a critical $p$ value the largest cluster size increases following a second-order transition. 
As evident from the plot, this increase is slower than in the case of the PR model or decreasing weight
order of H-PHN. In the latter case, the largest component grows slowly at early stages, and the transition
occurs at a much larger $p$ value.
The transition is much steeper than in the random addition case, and is much closer to an explosive percolation process.

\begin{figure}
\includegraphics*[width=0.40\textwidth]{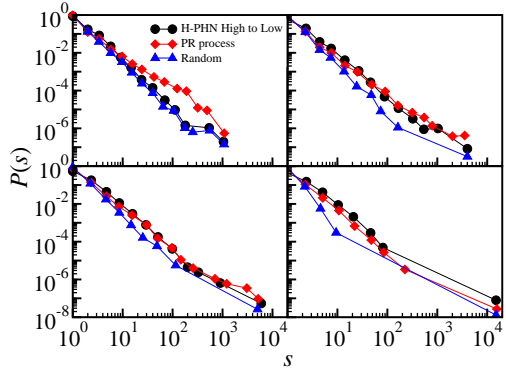}
\caption{Cluster size distribution, $P(s)$, for the PR model, random percolation, and the H-PHN with links added in decreasing order. The panels show four stages of $P(s)$ when the largest component size is 1000 nodes (upper left), 4000 nodes (upper right), 5000 nodes (lower left), 15000 nodes (lower right).
}
\label{component_size_dist}
\end{figure}

\begin{figure*}
\includegraphics*[width=1\textwidth]{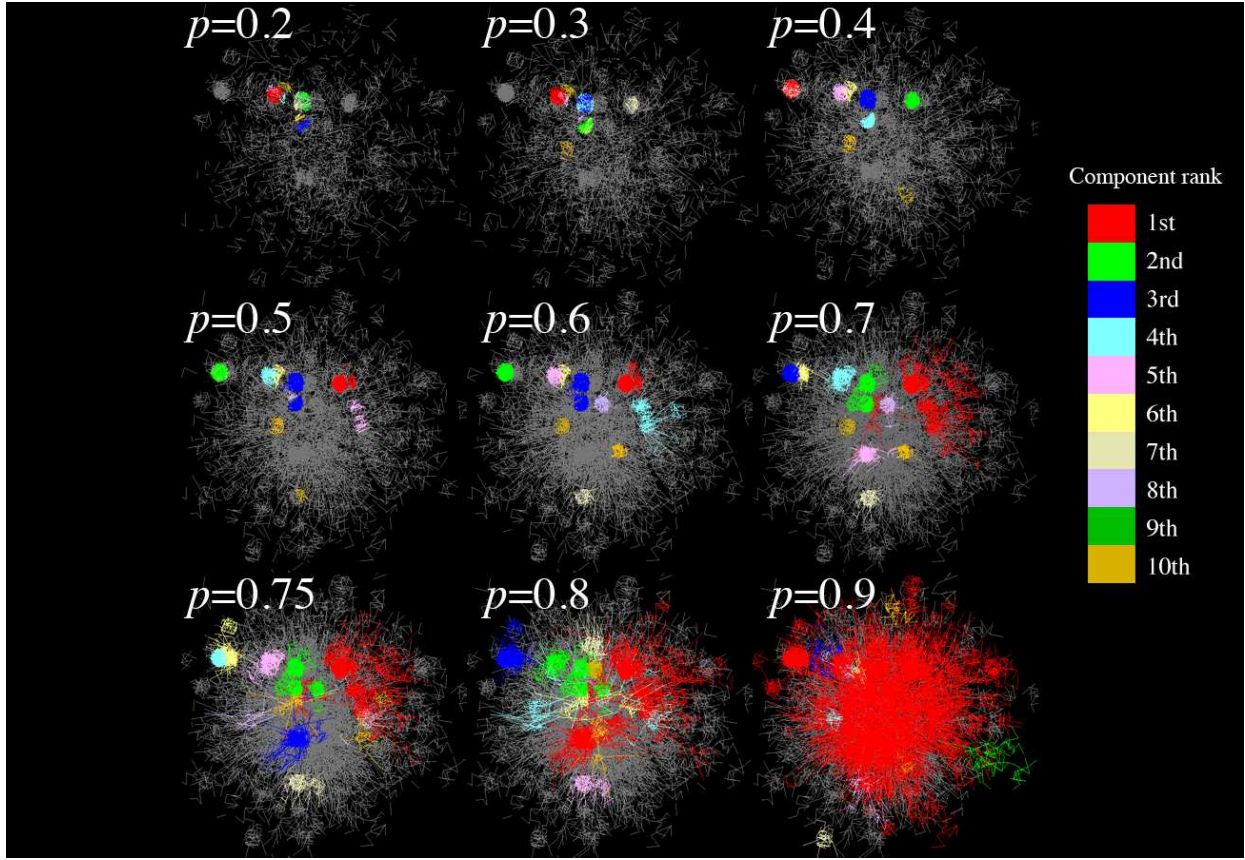}
\caption{Snapshots of the H-PHN network when different fraction $p$ of links have been added
according to their decreasing weight value. For each network we draw only the links that have been
added at this $p$ value (grey color), and we highlight the ten largest clusters with the colors
shown in the colormap.
}
\label{cool}
\end{figure*}

A clearer picture of how modules evolve during link addition emerges from Fig.~\ref{component_product}.
When we add a link, this link joins two clusters into one. Here, we calculate the product, $S_AS_B$, of the sizes
 of the two clusters that are joined through this link.
This product is then plotted as a function of the links fraction $p$ that have been added up to that point.
For the case of decreasing order of weights, this product remains very small until we have added roughly 80\% of
the network links. This shows that we only connect very small clusters with each other. Near the transition point,
though, this value increases abruptly and peaks at values much larger than the network size, showing that
we merge progressively larger clusters. Similar is the case for the explosive percolation, where the peak
is even more pronounced. In both cases, it is only at the late stages that the product $S_AS_B$ becomes
large and greatly exceeds the network size.
On the other hand, when links are added in increasing order the product $S_AS_B$ grows linearly with time until it reaches its
maximum at $S_AS_B=N$, where $N$ is the size of the network. This is consistent with the behavior observed
in Fig.~\ref{largest_component_phn}, where almost all new links are added to the largest cluster, so that 
with high probability $S_A\sim1$ and $S_B\sim p$, resulting in a linear increase.
When links are added at random, we observe the expected percolation transition at $p=0.5$, after which the giant component
grows rapidly and spans the entire network. The remaining links only attach small clusters to this cluster, so that its value
remains constant around $N$ from $p=0.5$ to $p=1$.

In Fig.~\ref{component_size_dist} we compare the cluster size distribution, $P(s)$, for the PR process, random percolation and the H-PHN. Below the percolation transition, when the size of the largest component is 1000 nodes (upper left panel of Fig.~\ref{component_size_dist}), $P(s)$ presents a similar behavior for all three processes, following roughly a power-law decay. Thus, clusters of all sizes (up to the largest one) appear in the network, with the PR model showing a slight preference towards larger clusters. As, though, we move closer to the percolation transition (see upper right and lower left panels of Fig.~\ref{component_size_dist}), the random process leads to a dominant largest cluster, with the remaining components being orders of magnitude smaller than the spanning cluster. In contrast, the distribution, $P(s)$, for the PR process and the H-PHN remains a power-law, indicating the existence of many clusters of significant size. When we consider $P(s)$ above the transition point (lower right panel of Fig.~\ref{component_size_dist}), the largest component dominates in all three processes, although the PR process and the H-PHN display a persistence of clusters of intermediate size. This behavior highlights the similarities in the explosive behavior, even at a microscopic level, between the H-PHN and the PR model.


\section{DISCUSSION}
\label{discussion}

Figure~\ref{cool} depicts the process of cluster formation and the corresponding connections between them as the value of
$SR$ is lowered, or equivalently as we add more links according to decreasing weight. The underlying network representation,
i.e. the spatial positions of each protein, is based on a Kamada-Kawaii~\cite{kamada} algorithm
on the final connected network (at $p=1$). Clearly, a large number of clusters are created very early in the process
and they retain their identity even for large values of $p$. For example, even when we have added 60\% of the total
links the ten largest clusters are more or less the same as the ones that were present at $p=0.2$. As we increase
$p$ even further a larger component starts to appear at roughly $p=0.8$, but even at $p=0.9$ we can clearly
detect isolated modules that are not yet attached to the spanning cluster. The largest component seems to grow
by absorbing dense clusters.

The process of adding links from high to low weights not only resembles the conditions for explosive percolation
but can also be considered analogous to the product rule presented in Ref.~\cite{achlioptas}.
In this paper we have shown that, on the average, when new links are added to the network they
connect clusters in a way that the product of their component sizes remains small,
and therefore the growth of the largest component is delayed. In the opposite case, when low weight links are added first, 
an accelerated growth of the largest component is observed.

Our results indicate that a modular network, such as H-PHN, can be well-described through the
idea of explosive percolation, although one cannot conclusively characterize the nature of the transition, since 
finite size scaling is not possible in real-world networks. 
It is important, though, to remark that a one-to-one correspondence between the H-PHN and the 
Achlioptas process is in principle impossible to determine:  The competition between links in an 
Achlioptas process is extremely hard (if possible) to detect in real-world networks because one has 
only indirect knowledge of the links preference to be established during network growth. For this reason 
we may only detect features of an Achlioptas process by studying the main properties leading to a sharp 
transition, and not the mechanism itself. In this sense, the competition between links in the H-PHN is 
inherent in the weights of links, which are determined using purely biological information. A strong 
indication of the similarity between the Achlioptas process and the H-PHN is shown in Fig.~\ref{component_product}, 
where the product of the components that are connected during the addition of links remains small 
for small values of $p$ and has a sharp peak (that greatly exceeds the size of the network) at a late stage in the process.

The sharp transition of the H-PHN is not a universal property ascribed to weighted networks. The inset of Fig.~\ref{largest_component_phn} shows that when the link weights of the H-PHN are shuffled without modifying the structure of the network, we observe a smooth percolation transition that appears at an early stage in the addition of links. Similarly, the same growth pattern emerges in another strongly modular network, the network of movie actors obtained from IMDB in which two actors are connected if they co-acted in a movie. The weight of a link corresponds to the number of movies in which they co-acted. Although the IMDB network displays a strong community structure and is organized in well-defined modules, there are many inter-module links at early stages, and therefore the giant component emerges continuously. In both cases the link weight distribution cancels out the effect of the modular substrate. Therefore, we find that sharp percolation transition is not necessarily inherent in networks with well-defined communities. Instead, the sharpness of the transition strongly depends on the way in which the network evolves during the addition of links.


\acknowledgments

We thank Thomas Rattei for providing the data for the H-PHN, and Diego Rybski for useful insight.
We acknowledge support from NSF grants.

\section*{APPENDIX: THE PROTEIN HOMOLOGY NETWORK}

In this work, we have used a biological network as an example of a highly modular
structure whose properties can be described through a product rule Achlioptas process.
The dataset on the Human Protein Homology Network (H-PHN) was obtained from SIMAP~\cite{SIMAP}.
This network is composed of proteins in the human cell, where two proteins are linked if they are homologous.
Of course, any two proteins have a certain degree of homology, so that we start from a fully connected
weighted network. The homology between two proteins is calculated by SIMAP through the $E$-value.
The $E$-value quantifies the level of statistically significant similarity between proteins
by finding the best possible alignment between them using the Smith-Waterman algorithm~\cite{smith}.
The $E$-value ranges from 0 to infinity, where 0 indicates a perfect alignment.
In this work we consider links between proteins with an $E$-value up to a cut-off of $ 10^{-10}$.
The resulting network comprises $N=21709$ nodes and $M=1289345$ links.
After we obtain this network, we use the alignment score for the links weight, which is
an equivalent, but slightly more accurate measure than the $E$-value. The value of the score $S$
is also provided by SIMAP, and it detects the degree of similarity between two optimally aligned proteins.
The actual genetic distance between proteins is quantified through the score ratio $SR$, which for two proteins
1 and 2 is defined as
\begin{eqnarray*}
s_{ij} = \frac{\rm alignment~score~between~i~and~j}{\rm score~of~self-aligning~protein~i}, \\
SR = \rm{min} (s_{12},s_{21})
\end{eqnarray*}
The self-alignment score in the denominator properly normalizes the score $s_{ij}$, so that
it takes into account e.g. the varying length between two otherwise similar proteins.
Notice that $0 \leq SR \leq 1$. $SR=1$ indicates a perfect alignment between the two proteins,
or in other words, a short genetic distance between the proteins.
The values of $SR$ for a protein pair in the network corresponds to the weight of the corresponding
link that we finally used in this work.

\end{document}